\documentclass[preprint,showpacs,preprintnumbers,prb]{revtex4-1}

\usepackage{graphicx}
\usepackage{dcolumn}
\usepackage{bm}

\begin{document}

\title{Contrasting pressure evolutions of $f$ electron 
hybridized states in CeRhIn$_5$ and YbNi$_3$Ga$_9$: 
an optical conductivity study}

\author{H.~Okamura}\altaffiliation[Electronic address: ]{ho@tokushima-u.ac.jp}
\affiliation{Graduate School of Advanced Technology and Science, 
Tokushima University, Tokushima 770-8506, Japan}
\author{A. Takigawa}
\author{T. Yamasaki}
\affiliation{Graduate School of Science, 
Kobe University, Kobe 657-8501, Japan} 
\author{E. D. Bauer}
\affiliation{Los Alamos National Laboratory, Los Alamos, 
NM 87545, USA}
\author{S. Ohara}
\affiliation{Graduate School of Engineering, 
Nagoya Institute of Technology, Nagoya 466-8585, Japan }
\author{Y. Ikemoto}
\author{T. Moriwaki}
\affiliation{Japan Synchrotron Radiation Research Institute, 
Sayo 679-5198, Japan}

\date{\today}

\begin{abstract}
Optical conductivity [$\sigma(\omega)$] of 
CeRhIn$_5$ and YbNi$_3$Ga$_9$ have been measured at 
external pressures to 10~GPa and at low 
temperatures to 6~K.  
Regarding CeRhIn$_5$, at ambient pressure 
the main feature in $\sigma(\omega)$ is a Drude peak 
due to free carriers.  With increasing pressure, however, 
a characteristic mid-infrared (mIR) peak rapidly 
develops in $\sigma(\omega)$, and its peak energy 
and width increase with pressure.    These features 
are consistent with an increased conduction ($c$)-$f$ 
electron hybridization at high pressure, and show 
that the pressure has tuned the electronic state of 
CeRhIn$_5$ from very weakly to strongly hybridized ones.  
As for YbNi$_3$Ga$_9$, in contrast, a marked mIR 
peak is observed already at ambient pressure, 
indicating a strong $c$-$f$ hybridization.  
At high pressures, however, the mIR peak shifts to 
lower energy and becomes diminished, and seems 
merged with the Drude component at 10~GPa.  
Namely, CeRhIn$_5$ and YbNi$_3$Ga$_9$ exhibit some 
opposite tendencies in the pressure evolutions 
of $\sigma(\omega)$ and electronic structures.    
These results are discussed in terms of the pressure 
evolutions of $c$-$f$ hybridized electronic states 
in Ce and Yb compounds, in particular in terms of 
the electron-hole symmetry often considered between 
Ce and Yb compounds.  
\end{abstract}


\maketitle
\section{Introduction}
Physics of strongly correlated $f$-electron systems, 
most typically Ce-based and Yb-based intermetallic 
compounds, has attracted much interest for the 
last few decades.\cite{onuki}   
Central to the problem is a duality between 
localized and delocalized characteristics 
exhibited by the $f$ electrons.  The $f$ electrons 
intrinsically exhibit localized characteristics 
since the $f$ orbitals are located closer to 
the nucleus than the conduction states.   
However, they may become partially delocalized 
by hybridizing with conduction ($c$) electrons.   
This $c$-$f$ hybridization leads to various 
interesting phenomena such as the Kondo effect, 
heavy fermion (HF) formation, intermediate valence 
(IV), Rudermann-Kittel-Kasuya-Yoshida 
interaction and the associated magnetic ordering.  
It also plays an important role in the quantum 
critical phenomena (QCP) at the border of magnetic 
ordering.

In IV compounds, the $c$-$f$ hybridization is 
fairly strong, and the average Ce or Yb valence 
significantly deviates from 3 and takes an 
intermediate value well above 
and below 3 for Ce and Yb compounds, 
respectively.\cite{lawrence,riseborough}  
Optical conductivity [$\sigma(\omega)$] studies 
have provided much information about their 
microscopic electronic states.\cite{wang}   
A marked mid-infrared (mIR) peak has been 
commonly observed in $\sigma(\omega)$ of many 
Ce- and Yb-based IV metals, and its origin has 
been discussed in terms of the $c$-$f$ 
hybridized electronic states.
\cite{sievers,garner,degiorgi,dordevic,hancock,
universal,pines,mutou,saso,kimura-ce,kimura-yb,
kimura-ce2,115-burch}  
For example, a model of ``renormalized $c$-$f$ 
hybridized bands'' has been used to understand the mIR 
peak.\cite{garner,degiorgi,dordevic,hancock,
universal}   
In this model, a flat $f$ band renormalized by large 
$f$ electron correlation ($U$) hybridizes with 
a wide $c$ band, forming a pair of hybridized 
bands near the Fermi level ($E_{\rm F}$).\cite{cox,coleman1,coleman2}   
The mIR peak in this model results from optical 
excitations between the two bands.\cite{garner}   
Its peak energy is given as 
$E_{\rm mIR} \simeq 2 \widetilde{V}$, where 
$\widetilde{V}$ is the $c$-$f$ hybridization 
renormalized by large $U$, and expressed as 
\begin{center}
\begin{equation}
\widetilde{V} \simeq \sqrt{T_K W},
\end{equation}
\end{center}
where $T_K$ and $W$ indicate the Kondo temperature 
and $c$ bandwidth, respectively.\cite{cox,coleman1,coleman2}   
Measured $E_{\rm mIR}$ values of different IV metals 
have been compared with their $\sqrt{T_{\rm K} W}$ 
(or related quantities) estimated by other experiments, 
and a universal relation between $E_{\rm mIR}$ and 
$\sqrt{T_{\rm K} W}$ has been found over a 
variety of Ce and Yb 
compounds.\cite{dordevic,hancock,universal,pines}   
An example of such universal relation\cite{universal} 
is shown in Fig.~1.  
\begin{figure}[t]
\begin{center}
\includegraphics[width=0.65\textwidth]{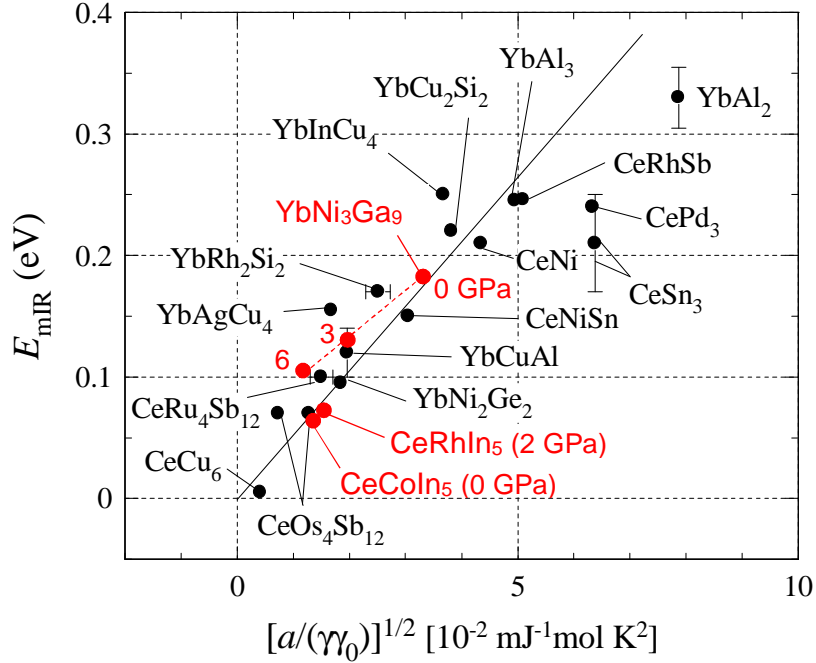}
\caption{(Color online) 
A universal relation between the optical conductivity 
and the $c$-$f$ hybridization energy observed for 
Ce and Yb compounds (the red data points have been 
newly added here, while the others have been 
reproduced from Ref.~\onlinecite{universal}).   
Here, the mid-IR peak energy ($E_{\rm mIR}$) measured 
for these compounds are plotted as a function of their 
$\sqrt{a/(\gamma \gamma_0)}$, where $\gamma$ and 
$\gamma_0$ are the specific heat coefficients of the 
Ce (Yb) and La (Lu) compounds, respectively, and $a$ 
is a $f$-degeneracy dependent constant.\cite{universal}  
Here, $\sqrt{a/(\gamma \gamma_0)}$ is a measure of 
the $c$-$f$ hybridizaton energy $\widetilde{V}$, 
through the relation $2 \widetilde{V} \simeq \sqrt{T_{\rm K}W} 
\propto \sqrt{(a/\gamma)(1/\gamma_0)}$.  
The solid line is guide to the eye.  
}
\end{center}
\end{figure}
%
This universal relation may be regarded as an 
optical analogue to the well known Kadowaki-Woods 
relation.\cite{KW,grand-KW}  
More detailed analyses including effects of $f$ 
level degeneracy and/or the band structure have 
also been reported.
\cite{mutou,saso,kimura-ce,kimura-yb,kimura-ce2}   
These studies suggest that the $c$-$f$ 
hybridized band model is an oversimplification 
for the actual IV metals.   
For example, it has been suggested that, for Ce 
compounds, $E_{\rm mIR}$ should correspond to 
the energy separation from the $c$-$f$ hybridized 
band below $E_{\rm F}$ to the bare 
$f$ states above $E_{\rm F}$.\cite{kimura-ce2}   
In fact, the observed $E_{\rm mIR}$ values of 
some IV compounds seem too large to result 
between the $c$-$f$ hybridized bands, and such a 
model may offer a useful alternative to 
the $c$-$f$ hybridized band model.  
Nevertheless, it is still true that the mIR peak 
energy is roughly scaled with $\sqrt{T_K W}$ over 
many IV metals.\cite{dordevic,hancock,universal,pines}  
Clearly, the characteristics of the mIR peak involve 
the Kondo physics, and are not due to accidental 
band structures.  
Furthermore, effects of momentum-dependent $c$-$f$ 
hybridization have been considered in analyzing 
$\sigma(\omega)$ of Ce compounds.\cite{115-burch}

Note that both Ce- and Yb-based IV metals seem to 
follow the same universal relation,\cite{universal,pines} 
as seen in Fig.~1.  For Ce$^{3+}$ and 
Yb$^{3+}$ ions, their respective $f^1$ and $f^{13}$ 
configurations have an electron-hole ($e$-$h$) 
symmetry, since $f^{13}$ is equivalent to $h^1$.   
It has been an important question as to what degree 
this $e$-$h$ symmetry is reflected on the 
properties of Ce and Yb compounds.   
An example of common property between them, which is 
consistent with the $e$-$h$ symmetry, is the formation 
of HF state with large effective mass.  
However, Ce and Yb compounds also exhibit noted 
differences.\cite{flouquet,122hikaku} 
A useful experimental technique to examine the 
$e$-$h$ symmetry is the application of an external 
pressure ($P$).\cite{flouquet,122hikaku,Yb-theory,thompson}  
Since Ce$^{4+}$ ($f^0$) and Yb$^{3+}$ ($f^{13}$) ions 
have smaller ionic radii than Ce$^{3+}$ ($f^1$) and 
Yb$^{2+}$ ($f^{14}$) ions, respectively, an applied $P$ 
generally increases the average valence ($v$) of Ce 
toward 4 and that of Yb toward 3.   
For both Ce and Yb cases, $P$ should also increase the 
bare (unrenormalized) $c$-$f$ hybridization, since a 
reduced interatomic distance should increase the overlap 
between the $c$ and $f$ wave functions.    
For Ce compounds, an increase of $c$-$f$ 
hybridization with $P$ has been observed, for 
example, by an increase in $T_{\rm K}$.\cite{flouquet}    
Then, in Fig.~1, upon applying $P$, a Ce compound 
should move to upper right. 
For Yb compounds, in addition to $v$ increases, 
effective mass increases and magnetic order have 
been found at high $P$.\cite{flouquet,122hikaku,thompson}  
Namely, Yb compounds seem to exhibit more localized $f$ 
electron states at high $P$.  This suggests a reduced 
$T_K$, and hence a reduced $\widetilde{V}$ from Eq.~(1), 
although the bare hybridization in a Yb compound should 
increase with $P$ as explained above.  Therefore, it 
is intriguing how an Yb compound should move with 
$P$ in Fig.~1.

In this work, we have addressed the above questions 
by studying the $\sigma(\omega)$ of CeRhIn$_5$ and 
YbNi$_3$Ga$_9$ at $P$ to 10~GPa and at temperatures 
($T$) to 6~K.   These 
compounds have attracted much attention for their 
remarkable properties at high $P$, as 
summarized in Fig.~2.  
%
\begin{figure}[t]
\begin{center}
\includegraphics[width=0.7\textwidth]{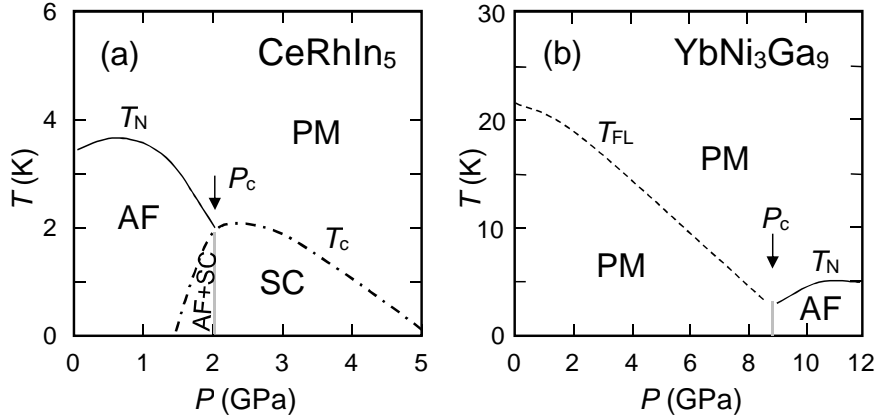}
\caption{
Schematic phase diagrams of (a) CeRhIn$_5$ 
(after Ref.~\onlinecite{115-phase2}) and (b) YbNi$_3$Ga$_9$ 
(after Ref.~\onlinecite{139-pressure})
as functions of temperature 
($T$) and external pressure ($P$).  
AF: antiferromagnetic, PM: paramagnetic, SC: superconducting, 
$T_N$: Neel temperature, $T_c$: superconducting transition 
temperature, $T_{FL}$: the temperature below which Fermi 
liquid characteristics are observed, $P_c$: critical pressure 
where AF ordering appears or disappears. 
} 
\end{center}
\end{figure}
%
CeRhIn$_5$ exhibits an antiferromagnetism (AF) at 
$P$=0 with a Neel temperature ($T_N$) of 3.3~K and 
an electronic specific heat coefficient of 
$\gamma$= 420~mJ/K$^2$mol above $T_N$.\cite{115-sc}  
With increasing $P$, the AF is gradually suppressed, 
and near a critical pressure ($P_c$) of $\sim$ 2~GPa, 
a superconductivity with a transition temperature 
($T_c$) of 2.1~K is observed.\cite{115-sc,115-dac}   
Around $P_c$, various anomalous properties related 
with QCP have been 
observed.\cite{115-review,115-onuki,115-phase,115-phase2}  
$\sigma(\omega)$ of CeRhIn$_5$ at ambient $P$ has 
already been measured and analyzed in 
detail,\cite{115-vdmarel,115-burch} but $\sigma(\omega)$ 
at high $P$ had not been explored yet.   
YbNi$_3$Ga$_9$, in contrast, is a paramagnetic IV 
compound at $P$=0 with $\gamma$=30~mJ/K$^2$mol, 
indicating a strong $c$-$f$ 
hybridization.\cite{139-ohara,139-PES}    
With increasing $P$, the measured $v$ increases 
from 2.6 at $P$=0 to 2.88 at $P$=16~GPa, and an AF 
state appears above $P_c \simeq$ 9~GPa.\cite{139-pressure}  
In addition, $\gamma$ increases significantly 
with $P$, reaching $\gamma$=1~J/K$^2$mol at 
9~GPa.\cite{139-gamma}  
Namely, with increasing $P$, CeRhIn$_5$ shows a 
crossover from localized to delocalized electronic 
states, while YbNi$_3$Ga$_9$ shows that from 
delocalized to localized ones.   
Although the lowest $T$'s in our study, 
6~K for CeRhIn$_5$ and 8~K for YbNi$_3$Ga$_9$, are 
above $T_c$ and $T_N$, our study should still provide 
important information about the $P$ tuning of the 
underlying $c$-$f$ hybridized state behind the 
QCP-related properties below $T_c$ and $T_N$.  
The mIR peaks of CeRhIn$_5$ and YbNi$_3$Ga$_9$ 
have indeed shown quite contrasting $P$ evolutions, 
which are discussed in terms of the $c$-$f$ 
hybridized electronic states at high $P$, and 
in terms of the $e$-$h$ symmetry.  

\section{Experimental}
The samples of CeRhIn$_5$ and YbNi$_3$Ga$_9$ used were 
single crystals grown with self-flux method.  
The reflectance spectrum [$R(\omega)$] of a sample was 
measured on an as-grown surface without polishing.  
$\sigma(\omega)$ was derived from $R(\omega)$ using the 
Kramres-Kronig (KK) analysis.\cite{dressel}  
$R(\omega)$ at $P$=0 was measured at photon energies 
between 15~meV and 30~eV covered by several light 
sources,\cite{okamura-chapter} 
including the vacuum uv synchrotron radiation 
at the beamline BL7B of the UVSOR Facility.\cite{bl7b}    
$R(\omega)$ spectra at high $P$ were measured using 
a diamond anvil cell (DAC).\cite{pressure-review}  
Type IIa diamond anvils with 0.8~mm culet diameter 
and a stainless steel gasket were used to seal the 
sample with glycerin as the pressure transmitting 
medium.\cite{medium1,medium2,medium3}   
A flat surface of a sample was closely attached 
on the culet surface of the diamond anvil, and 
$R(\omega)$ at the sample/diamond interface was measured.   
Small ruby pieces were also sealed to monitor the 
pressure via its fluorescence.   
A gold film was placed between the gasket and 
anvil as a reference of $R(\omega)$.     
In the KK analysis of $R(\omega)$ measured with DAC, 
the refractive index of diamond ($n_d$=2.4) was taken 
into account as previously discussed.\cite{kk-dia}
$R(\omega)$ at high $P$ and low $T$ were measured at 
photon energies from 25~meV (CeRhIn$_5$) or 20~meV 
(YbNi$_3$Ga$_9$) to 1.1~eV, using synchrotron 
radiation as a bright IR source\cite{JPSJ-review} at 
the beamline BL43IR of SPring-8.\cite{micro1,micro2}  
Below the measured energy range, $R(\omega)$ was 
extrapolated with the Hagen-Rubens 
function.\cite{dressel}  
More details of the high pressure IR experiments 
can be found elsewhere.\cite{pressure-review}

\section{Results and Discussions}
\subsection{$R(\omega)$ and $\sigma(\omega)$ of CeRhIn$_5$ 
at high pressures}
Figure~3(a) shows $R(\omega)$ and $\sigma(\omega)$ of 
CeRhIn$_5$ at $P$=0 and $T$=295~K over the entire 
measured spectral range.  $R(\omega)$ below 0.3~eV is 
very high, which indicates highly metallic characteristics 
of CeRhIn$_5$.   $\sigma(\omega)$ has a Drude component 
rising toward zero energy, which is due to free carrier 
dynamics.   
Figures~3(b) and 3(c) show $R(\omega)$ and $\sigma(\omega)$ 
measured at different values of $P$ and $T$.     
%
\begin{figure}
\begin{center}
\includegraphics[width=0.9\textwidth,clip]{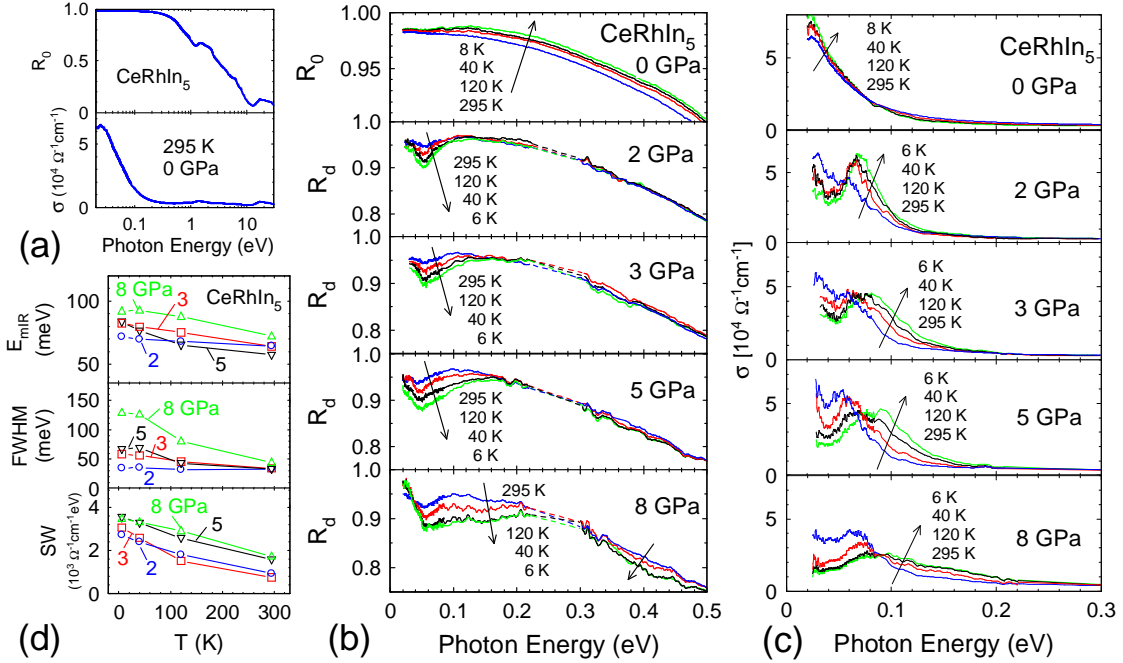}
\caption{(Color online) 
(a) $R(\omega)$ and $\sigma(\omega)$ spectra of 
CeRhIn$_5$ measured at $T$=295~K and $P$=0.  
(b) $R(\omega)$ at $P$=0, 2, 3, 5, and 8~GPa 
and at $T$=295~K (blue curves), 120~K(red), 
40~K (black), and 8~K (0~GPa) or 6~K (2-8~GPa) 
(green).  
$R_0$ and $R_d$ denote $R(\omega)$ measured at 
sample/vacuum and sample/diamond interfaces, 
respectively.  
The broken-curve portions of the spectra indicate 
interpolations, which were needed because of strong 
absorption by the diamond.\cite{comment5}  
(c) $\sigma(\omega)$ at the same values of 
$P$ and $T$ as in (b).  
(d) Peak position ($E_{\rm mIR}$), the full width at 
half maximum (FWHM), and the spectral weight (SW) of 
the mIR peak in $\sigma(\omega)$, given by the spectral 
fitting.   
}
\end{center}
\end{figure}
%
At $P$=0, $R(\omega)$ and $\sigma(\omega)$ have only 
minor $T$ dependences, which is consistent with the 
previous report.\cite{115-vdmarel}  
With increasing $P$, as shown in Figs.~3(b) and 3(c), 
a dip appears and develops in $R(\omega)$, and a 
pronounced mIR peak develops in $\sigma(\omega)$.  
At 2~GPa, the mIR peak is barely visible at 295~K, 
but becomes very pronounced with cooling to 6~K.  
As discussed in Introduction, such a mIR peak is 
a hallmark of the $c$-$f$ hybridized state in 
Ce compounds.   
Namely, CeRhIn$_5$ at 2~GPa has much stronger $c$-$f$ 
hybridization than at 0~GPa.  
Note that the $T$-evolution of mIR peak at $P$=2~GPa 
is strikingly similar to that of CeCoIn$_5$ at $P$=0, 
with very close $E_{\rm mIR}$.\cite{basov,115-vdmarel,okamura115}  
CeCoIn$_5$ at $P$=0 is also a superconductor with almost 
the same $T_c$ as that of CeRhIn$_5$ at 2~GPa.\cite{basov}   
These similarities between CeCoIn$_5$ at 0~GPa and CeRhIn$_5$ 
at 2~GPa indicate that their $c$-$f$ hybridized electronic 
states are also similar.  
In Fig.~1, $E_{\rm mIR}$ of CeCoIn$_5$ at 0~GPa and 
CeRhIn$_5$ at 2~GPa have been added using 
their $\gamma$ and $\gamma_0$ data.\cite{comment2} 
Clearly, they follow the universal relation well, 
and their plots are indeed close to each other 
reflecting their similarity.  
At $P$=3~GPa, the mIR peak of CeRhIn$_5$ is broader 
than that at 2~GPa.  
This broadening of the mIR peak should basically 
indicate a broadening of the $f$ band, and hence a 
stronger $c$-$f$ hybridization.  
From the evolutions of $\sigma(\omega)$ from $P$=0 to 3~GPa, 
it is clear that the electronic structure of CeRhIn$_5$ in 
the normal state at 6~K changes significantly from very 
weakly to moderately hybridized ones, which should 
be an important basis for the QCP observed below 2~K.   
From 3 to 8~GPa, the mIR peak becomes apparently much broader, 
and its spectral weight (SW) shifts toward higher energy.  In 
addition, at 5 and 8~GPa the mIR peak is clearly observed even 
at room $T$, which is a feature often observed for IV Ce 
compounds.   Namely, CeRhIn$_5$ above 5~GPa is likely 
a strongly hybridized IV compound.  
Unfortunately, $\gamma$ data of CeRhIn$_5$ above 2~GPa 
are unavailable, so $E_{\rm mIR}$ above 2~GPa cannot 
be plotted in Fig.~1.  However, it is almost certain that 
the plot for CeRhIn$_5$ moves to upper right with $P$, 
since the resistivity data\cite{115-dac} strongly suggest 
that the hybridization is much stronger at 8~GPa.

To analyze the evolution of mIR peak more quantitatively, 
we have performed spectral fitting on the measured 
$\sigma(\omega)$ using the Drude-Lorentz oscillator 
model.\cite{dressel}   
Details of the fitting procedures and examples of the 
fitted spectra are given in the Appendix.  In Fig.~3(d), 
the $E_{\rm mIR}$, full width at half maximum (FWHM), 
and the SW of the mIR peak given 
by the fitting are summarized.  
The fitting results in Fig.~3(d) confirm the features 
discussed above, namely 
the $P$-induced increases in $E_{\rm mIR}$, SW, and the 
width.  However, they also reveal that the increases in 
$E_{\rm mIR}$ and SW are at most about 30~\%.  In contrast, 
the increase in the width at 6~K is particularly large 
from 5 to 8~GPa, suggesting that $f$ electron bandwidth 
should rapidly increase at this $P$ range.

\subsection{$R(\omega)$ and $\sigma(\omega)$ of 
YbNi$_3$Ga$_9$ at high pressures}
Figure~4(a) shows measured $R(\omega)$ and $\sigma(\omega)$ 
of YbNi$_3$Ga$_9$ over a wide energy range at $P$=0 and 
$T$=295~K, and Figures~4(b) and 4(c) show those below 0.4~eV 
at different values of $P$ and $T$.   
%
\begin{figure}[t]
\begin{center}
\includegraphics[width=0.85\textwidth,clip]{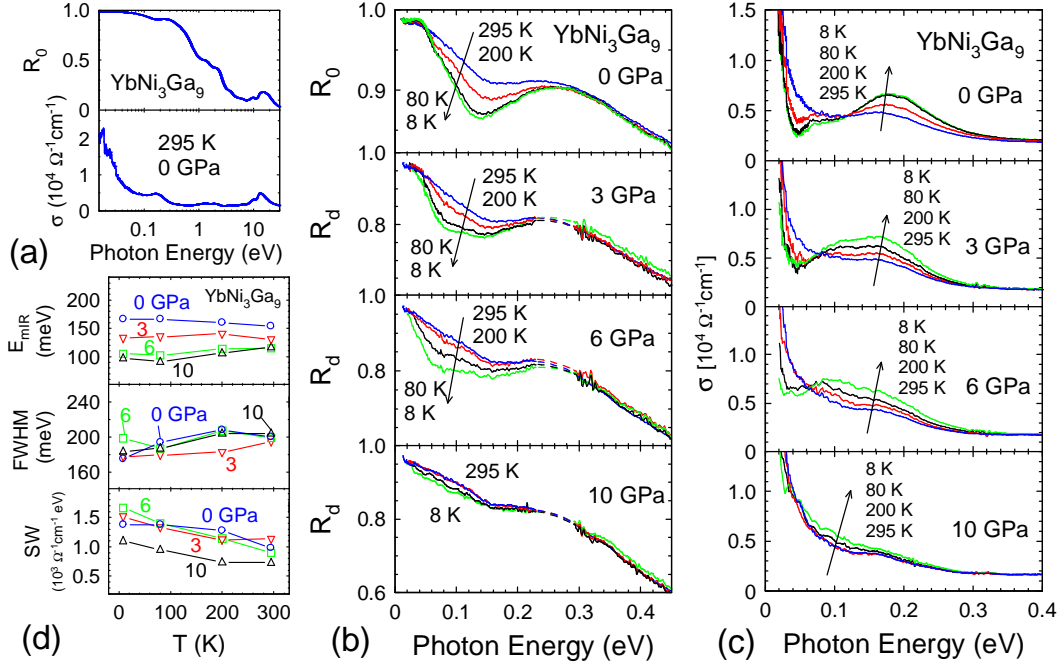}
\caption{(Color online) 
(a) $R(\omega)$ and $\sigma(\omega)$ spectra of 
YbNi$_3$Ga$_9$ measured at $T$=295~K and $P$=0.  
(b) $R(\omega)$ spectra at $P$=0, 3, 6, and 10~GPa 
and at $T$=295~K (blue curves), 200~K (red), 
80~K (black) and 8~K (green).  
The broken curve portions indicate smooth 
interpolations, as already mentioned in the 
caption of Fig.~3(b).\cite{comment5}  
(c) $\sigma(\omega)$ spectra at the same values 
of $P$ and $T$ as those in (b).  
(d) Results of the spectral fitting, with the 
same notations as those in Fig.~3(d).  
}
\end{center}
\end{figure}
%
At $P$=0, a dip is observed in $R(\omega)$ at 
0.1-0.2~eV range and a mIR peak in 
$\sigma(\omega)$ at 0.1-0.3~eV range, both of which 
are strongly $T$ dependent.  
A Drude component is also observed in $\sigma(\omega)$ 
below 0.1~eV, which rises steeply toward zero energy.    
With cooling, the dip becomes deeper, and the mIR peak 
becomes more pronounced and slightly shifts to higher energy.  
In addition, a shoulder appears in $\sigma(\omega)$ near 
70~meV at low $T$.  Furthermore, the onset of Drude 
component becomes extremely sharp at low $T$.  
This is because $\sigma(0)$ at low $T$ is very large, 
exceeding 1~$\times$ 10$^6$~$\Omega^{-1}$cm$^{-1}$ 
at 8~K.\cite{139-ohara,139-pressure}   
This extremely narrow Drude component is due to the 
Drude response of heavy quasiparticles formed 
at low $T$.\cite{sievers}  
These features are qualitatively very similar to 
those previously observed for other Yb-based 
IV metals such as YbAl$_3$.\cite{ybal3} 
Using the observed $E_{\rm mIR}$=0.18~eV at 8~K with 
$\gamma$=30~mJ/K$^2$mol and 
$\gamma_0$=6.3~mJ/K$^2$mol,\cite{139-ohara} a plot for 
YbNi$_3$Ga$_9$ at $P$=0 has been added in Fig.~1.  
Clearly, YbNi$_3$Ga$_9$ well follows the universal 
relation.

The $\sigma(\omega)$ spectra at various values of 
$T$ and $P$ have been analyzed by spectral fitting, 
similarly to those of CeRhIn$_5$.  
The results are summarized in Fig.~4(d), and examples 
of the fitting are given in Appendix.   
At 3~GPa, $R(\omega)$ and $\sigma(\omega)$ spectra are 
still strongly $T$ dependent.   Note that, at low $T$, 
the mIR peak seems to consist of two peaks, located at 
$\sim$ 0.1 and $\sim$ 0.17~eV.  A similar two-peak feature 
is also observed at 6~GPa.   The origins for the 
two peaks are unclear, hence we define $E_{\rm mIR}$ as 
the center-of-mass position of the Lorentz oscillators 
used to fit the mIR peak.    As indicated in Fig.~4(d), 
the obtained $E_{\rm mIR}$ decreases with $P$ from 
0 to 3~GPa, and also from 3 to 6~GPa.   
The $E_{\rm mIR}$'s at $P$=3 and 6~GPa and $T$=8~K 
given by the spectral fitting have been added in 
Fig.~1, using the $\gamma$ data measured at high 
$P$.\cite{139-gamma,comment3}     
The plot for YbNi$_3$Ga$_9$ moves to down left with 
$P$, namely, it actually moves in an opposite manner 
to that of Ce compounds.   
At 10~GPa, the mIR peak seems much weaker than at 6~GPa, 
and almost merged with the Drude component.  
The remaining mIR component has been evaluated by the 
fitting as in Fig.~4(d), which indicates that 
$E_{\rm mIR}$ further decreases compared with that at 6~GPa.  
From Figs.~4(b) and 4(c), the $T$ variations of 
$R(\omega)$ and $\sigma(\omega)$ at 10~GPa are 
much smaller than those at 6~GPa.  
This strongly suggests that the $T$-dependent hybridization 
has become much weaker with $P$ from 6 to 10~GPa.  
This is reasonable since the $f$ electron state 
should be more localized and less hybridized above 
$P_c \sim$ 9~GPa, where an AF state appears below 
$T_{\rm N}$=5~K.  
The fitting results in Fig.~4(d) also indicate that 
$E_{\rm mIR}$ does not change much with $T$, although 
it clearly decreases with $P$.  In addition, the peak 
width seems to exhibit no systematic changes with $T$ 
and $P$.

\subsection{Comparison of pressure evolutions 
between CeRhIn$_5$ and YbNi$_3$Ga$_9$ }  
To compare more clearly the observed $P$ evolutions of 
$\sigma(\omega)$ between CeRhIn$_5$ and YbNi$_3$Ga$_9$, 
the $\sigma(\omega)$ spectra at the lowest measured $T$ 
are displayed in Fig.~5.  
%
\begin{figure}[t]
\begin{center}
\includegraphics[width=0.65\textwidth,clip]{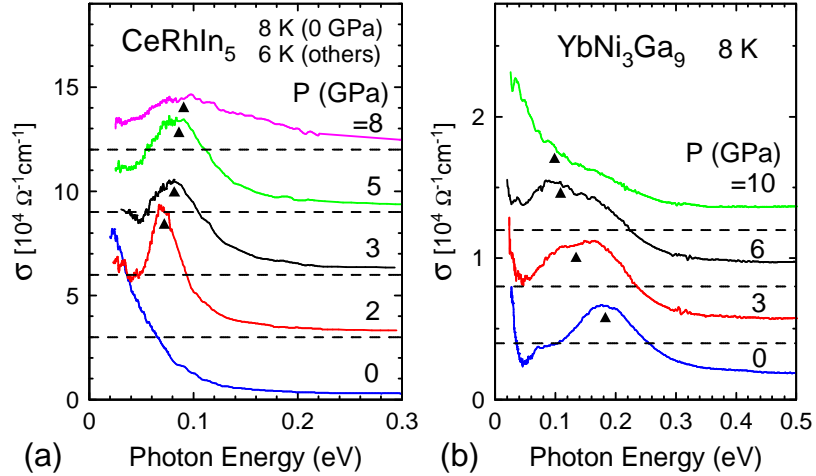}
\caption{(Color online) 
Comparison of $P$ evolutions of $\sigma(\omega)$ at 
low $T$ between (a) CeRhIn$_5$ and (b) YbNi$_3$Ga$_9$.  
The spectra are vertically offset for clarity, and 
the triangles indicate the $E_{\rm mIR}$ values.   
}
\end{center}
\end{figure}
%
Clearly, the $P$ evolutions of CeRhIn$_5$ 
and YbNi$_3$Ga$_9$ have some contrasting and opposite 
tendencies: With increasing $P$, the mIR peak of 
CeRhIn$_5$ appears and grows, and shifts to higher 
energy.  In contrast, the mIR peak of YbNi$_3$Ga$_9$ 
is well developed already at $P$=0, and shifts to 
lower energy with $P$ and becomes diminished at 10~GPa. 
On the other hand, not all the $P$ evolutions of mIR 
peak exhibit opposite tendencies between them.  For 
example, the mIR peak width of CeRhIn$_5$ significantly 
increases with $P$, but that of YbNi$_3$Ga$_9$ does 
not exhibit a narrowing or any other systematic change, 
as indicated in Fig.~6(b). 
In addition, $P$-induced shift of $E_{\rm mIR}$ seems 
much larger for YbNi$_3$Ga$_9$ than for CeRhIn$_5$.   
Below, we consider these results in terms of the 
$P$ evolution of IV states in Ce and Yb 
compounds.

In a Ce compound with $v$=3 and completely 
localized $f^1$ state, $\sigma(\omega)$ should 
consist only of a Drude component due to $c$ 
electrons, since the system is a metal where 
the localized $f$ electrons do not contribute 
to the Drude response.   
This applies well to CeRhIn$_5$ at 0~GPa, since its 
$\sigma(\omega)$ has only a Drude component in 
Fig~3(c) and its $v$ should be very close to 3.  
As stated in Introduction, an external $P$ on a 
Ce compound should increase $v$ from 3.  
In this IV state, $\sigma(\omega)$ would 
exhibit a mIR peak due to the hybridized state, 
as actually observed in $\sigma(\omega)$ of 
CeRhIn$_5$ at 2 and 3~GPa.  
With further increasing $P$, both the energy and 
width of the mIR peak should increase, since the 
hybridization and $f$ bandwidth increase with $P$.  
This is again consistent with the observed 
$\sigma(\omega)$ at 3-8~GPa.  
Namely, the observed $P$ evolutions of 
$\sigma(\omega)$ for CeRhIn$_5$ seem quite 
consistent with those expected for a Ce compound.    
$P$-induced higher-energy shifts and development 
of an IR peak in $\sigma(\omega)$ have also 
been observed for CeRu$_4$Sb$_{12}$ 
(Ref.~\onlinecite{okamura-sku}) 
and CeIn$_3$.\cite{iizuka}

As for Yb compounds, as stated in Introduction, 
an ionic radius consideration suggests that $v$ 
of an IV Yb compound should increase toward 3 with $P$.   
In the limit of exactly $v$=3 state with completely 
localized $f^{13}$ state, $\sigma(\omega)$ would 
consist only of a broad Drude component due to 
$c$ electrons, similarly to the $f^1$ (Ce$^{3+}$) 
case.  
Namely, the main feature in $\sigma(\omega)$ of 
an IV Yb compound should evolve from a mIR peak 
at $P$=0 into a broad Drude component in the limit 
of very high $P$.  
Clearly, such a $P$ evolution is qualitatively consistent 
with that observed for YbNi$_3$Ga$_9$ in Fig.~5(b): 
$\sigma(\omega)$ has a well developed mIR peak at 
$P$=0, which shifts to lower energy with $P$ and 
becomes much weaker at 10~GPa. 
$\sigma(\omega)$ at 10~GPa actually looks like a broad 
Drude component, and the residual mIR peak SW would 
become even weaker if $P$ is further increased 
since $v$ still increases from 2.84 at 10~GPa to 2.88 
at 16~GPa.\cite{139-pressure}

The discussions above indicate that the opposite 
tendencies in the $P$ evolutions of 
$\sigma(\omega)$ between CeRhIn$_5$ and YbNi$_3$Ga$_9$, 
including the opposite $P$-induced moves in Fig.~1, 
are consistent with the consideration of ionic radius 
and $e$-$h$ symmetry under high $P$.  
The expression for the renormalized hybridization 
$\widetilde{V}$ in Eq.~(1) has been derived for an 
$f^1$ (Ce$^{3+}$) system,\cite{cox,coleman1,coleman2} 
but is also valid for $h^1$ ($f^{13}$, Yb$^{3+}$) 
system under the $e$-$h$ symmetry.   Hence, for both 
Ce and Yb compounds, $\widetilde{V}$ and $E_{\rm mIR}$ 
should become smaller with increasing localization and 
decreasing $T_{\rm K}$.   This has been actually 
demonstrated by the Ce and Yb compounds plotted in Fig.~1.  
Then, according to Eq.~(1), the $P$-induced decrease 
of $E_{\rm mIR}$ for YbNi$_3$Ga$_9$ indicates that 
$\widetilde{V}$ decreases with $P$, although the bare 
(unrenormalized) hybridization should increase with 
$P$ as already discussed.    
This peculiar property of an Yb compound has been 
discussed\cite{mito} in terms of the $c$-$f$ 
exchange energy, $J_{cf}$, expressed as\cite{Jcf} 
\begin{equation}
J_{cf} \simeq  \frac{|V|^2}{|E_{\rm F} - \epsilon_f|},
\end{equation}
where $V$ is the bare $c$-$f$ hybridization 
averaged over the $k$ space and $\epsilon_f$ is 
the one-electron (unrenormalized) $f$ level.   
Eq.~(2) has been derived for an $f^1$ system 
with sufficiently large $U$, but is also valid 
for $h^1$ system under the $e$-$h$ symmetry.   
$J_{cf}$ is related with $T_{\rm K}$ as 
$T_{\rm K} \simeq W \exp{(-W/J_{cf})}$.  
Note that $|V|$ in Eq.~(2) increases with 
$P$ for both Ce and Yb cases, as discussed earlier.   
In addition, note that $\epsilon_f$ should increase 
with $P$ relative to $E_{\rm F}$.\cite{mito,comment9}   
For Ce case, $\epsilon_f$ is located below $E_{\rm F}$ 
and approaches $E_{\rm F}$ with increasing $P$.   
Hence $| E_{\rm F} - \epsilon_f |$ decreases in Eq.~(2), 
so that $J_{cf}$ increases with $P$.  
For Yb case, in contrast, $\epsilon_f$ is the $f$ hole 
level located above $E_{\rm F}$, and moves away from 
$E_{\rm F}$ with $P$.  Hence $| E_{\rm F} - \epsilon_f |$ 
increases with $P$ in Eq.~(2), so that $J_{cf}$ may 
either increase or decrease depending on which of 
$| E_{\rm F} - \epsilon_f |$ and $|V|^2$ increases more.  
Therefore, the $P$-induced decrease of $E_{\rm mIR}$ 
for YbNi$_3$Ga$_9$ suggests that $| E_{\rm F} - \epsilon_f |$ 
increases with $P$ more than $|V|^2$ does.

As already mentioned, some of the observed $P$ 
evolutions of $\sigma(\omega)$ are not opposite or 
symmetrical between CeRhIn$_5$ and YbNi$_3$Ga$_9$.  
For example, the mIR peak of CeRhIn$_5$ shows 
progressive and significant broadenings with $P$, 
while that of YbNi$_3$Ga$_9$ does not show a narrowing 
or any systematic change with $P$.  
In addition, $P$-induced shift of $E_{\rm mIR}$ 
seems much larger for YbNi$_3$Ga$_9$ (0.18 to 0.1~eV 
on going from 0 to 6~GPa) than for CeRhIn$_5$ 
(70 to 90~meV on going from 2 to 8~GPa). 
Microscopic mechanisms for these results are unclear, 
but they should involve microscopic differences 
between Ce$^{3+}$ and Yb$^{3+}$ not considered in 
the simple $e$-$h$ symmetry argument.  
For example, the 4$f$ orbital of Yb$^{3+}$ is much 
more localized than that of Ce$^{3+}$, leading to a 
much smaller $f$ bandwidth and $|V|$ for 
Yb$^{3+}$.\cite{flouquet,122hikaku}  
In addition, the spin-orbit splitting of Yb$^{3+}$ 
($\sim$ 1.3~eV) is much larger than that of Ce$^{3+}$ 
($\sim$ 0.3~eV).   It has been pointed 
out\cite{flouquet,122hikaku} that, due to these 
differences, the $P$-induced variation of $v$ from 
3.0 in a Ce compound should be at most to $\sim$ 
3.16, while that in an Yb compound can be changed 
more widely between 2 and 3.  
Experimentally, $v$ of CeRhIn$_5$ at high $P$ 
has not been reported, but that of CeCoIn$_5$ has 
been reported to vary from 3.00 at $P$=0 to 3.05 at 
8~GPa.\cite{yamaoka}  
On the other hand, $v$ of YbNi$_3$Ga$_9$ varies 
from 2.60 at $P$=0 to 2.84 at 
$P$=10~GPa.\cite{139-pressure}  
These different ranges of variation in $v$ may 
be related to the much larger $P$-induced 
shifts of $E_{\rm mIR}$ for YbNi$_3$Ga$_9$, since 
$v$ is closely related with $T_{\rm K}$ and 
$\widetilde{V}$.    
To further understand $P$ evolutions of $\sigma(\omega)$ 
and electronic structures for Yb compounds, 
more studies on other Yb-based IV compounds are 
clearly needed.   
For example, YbCu$_2$Ge$_2$ (Ref.~\onlinecite{miyake}) 
and YbAl$_2$ (Ref.~\onlinecite{dallera}) are other 
examples that exhibit large $P$ dependences in their 
physical properties.  $\sigma(\omega)$ studies of these 
compounds at high $P$ are in progress.\cite{Yb122}

\section{Summary}
$\sigma(\omega)$ studies of CeRhIn$_5$ and 
YbNi$_3$Ga$_9$ at high $P$ have been performed 
to probe the $P$ evolutions of their $c$-$f$ 
hybridized electronic structures. 
The main feature in the measured $\sigma(\omega)$ 
is a mIR peak, which has exhibited many opposite 
or symmetrical $P$ evolutions between CeRhIn$_5$ 
and YbNi$_3$Ga$_9$: With increasing $P$, the 
mIR peak develops and shifts to higher energy 
for CeRhIn$_5$, while it shifts to lower energy 
and becomes diminished at high $P$ for YbNi$_3$Ga$_9$.   
These results are qualitatively consistent with 
the $e$-$h$ symmetry and $P$-induced variations 
in the Ce and Yb ionic radii.  
However, YbNi$_3$Ga$_9$ has also exhibited $P$ 
evolutions of mIR peak not opposite to those of 
CeRhIn$_5$, which are likely due to microscopic 
differences between Ce and Yb not included in 
the simple $e$-$h$ symmetry arguments.

\begin{acknowledgments}
H. O. would like to thank Dr. Takeshi Mito and 
Dr. Tetsuya Mutou for useful discussions.  
The experiments at SPring-8 were performed under 
the approval by JASRI (2011B0089, 2012A0089, 2012B0089, 
2013A0089, 2013B0089, 2013B1159), and those at UVSOR 
under the approval by Institute for Molecular Science.  
H.~O. acknowledges financial support from JSPS KAKENHI 
(21102512, 23540409, 26400358).    
\end{acknowledgments}

\appendix* \section{Spectral fittings on $\sigma(\omega)$}
The spectral fittings were performed using the 
Drude-Lorentz model.\cite{dressel}   In this model, 
the complex dielectric function is expressed as a 
sum of Drude and Lorentz oscillators, which represent 
free and bound electrons, respectively, as
\begin{equation}
\hat{\epsilon}(\omega)=\epsilon_\infty + 
\sum_{j} \frac{\omega_{p,j}^2}
{\omega_{0,j}^2 - \omega^2 - i \omega \gamma_j}. 
\label{eq:DL}
\end{equation}
Here, $\omega_p$, $\omega_0$, and $\gamma$ are the 
plasma frequency, natural frequency, and damping, 
respectively.  $j$ denotes the $j$th oscillator, 
and $\omega_0$=0 for a Drude oscillator.  
$\varepsilon_\infty$ represents contribution 
from higher-lying interband transitions.   
In the fitting, these parameters are adjusted so 
as to reproduce a measured $\sigma(\omega)$ through 
the relation 
$\sigma(\omega)=(\omega/4\pi)\mbox{Im}(\tilde{\epsilon})$.

Figure~6(a) shows examples of fitting for 
$\sigma(\omega)$ of CeRhIn$_5$ at 6~K.  
%
\begin{figure}
\begin{center}
\includegraphics[width=0.65\textwidth,clip]{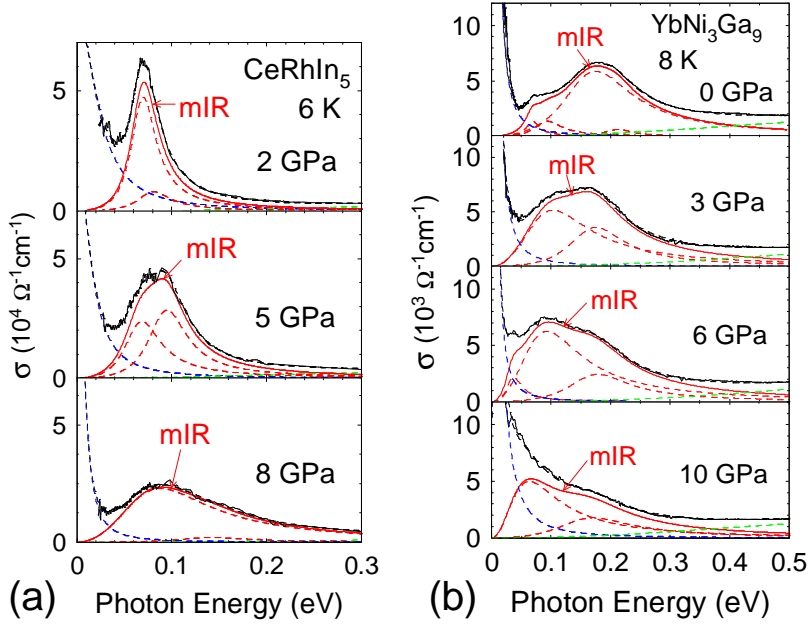}
\caption{(Color online) 
(a) Examples of Drude-Lorentz fitting on $\sigma(\omega)$ 
of CeRhIn$_5$ measured at $T$=6~K and at $P$=2, 5, and 
8~GPa, and (b) those of YbNi$_3$Ga$_9$  measured at 8~K at 
$P$=0, 3, 6, and 10~GPa, as discussed in the Appendix.   
Shown in each graph are the measured data (black solid 
curves), total fit (black broken), Drude component (blue 
broken), two Lorentz oscillators (red broken), mIR peak 
(red solid) which is the sum of the Lorentz oscillators, 
and the background (green broken). }
\end{center}
\end{figure}
%
The measured $\sigma(\omega)$ spectra can be reproduced 
well by using two Lorentz oscillators for the mIR peak, 
in addition to a Drude oscillator and a broad Lorentz 
oscillator peaked at 0.45~eV 
serving as a background.   $\epsilon_\infty$=5 
is used for all the fitting, and the fitted mIR peak is 
the sum of the two Lorentz oscillators.   
Although not shown, $\sigma(\omega)$ at other values 
of $P$ and $T$ can be fitted similarly.  
Note that two Lorentz oscillators are used simply because 
a single Lorentz oscillator is not sufficient to fit the 
mIR peak, and that the two Lorentz oscillators are not 
assigned to any specific origins.  
In addition, due to the use of DAC, the measured 
spectral range does 
not cover low-enough energies for fitting the Drude component.  
However, since our main focus here is the evolution of 
mIR peak with $P$ and $T$, the uncertainty regarding the 
Drude peak fitting is not a serious problem.  
To reduce the uncertainty in fitting the Drude component, 
$\sigma(0)$ value given by the fitting was kept in the 
range 8-20~$\times$ 10$^{4}$~$\Omega^{-1}$cm$^{-1}$, and 
$\sigma(0)$ was increased with decreasing $T$.  These 
constraints on the Drude component are implied from the 
$\sigma(0)$ at $P$=0 [Fig.~3(c)] and also from the 
measured dc conductivities.\cite{115-sc}

Figure~6(b) shows examples of fitting for YbNi$_3$Ga$_9$.  
In some cases more than two Lorentz oscillators 
are needed to fit the mIR peak reasonably well.   
A background Lorentz oscillator at 0.75~eV and 
$\epsilon_\infty$=5 are used.  
Here, the fitting parameters are chosen so that 
$\sigma(0)$ at each data matched the measured dc 
conductivity at the same $T$ and 
$P$.\cite{139-ohara,139-pressure,comment1}  
This procedure greatly reduced the uncertainty 
in fitting the Drude component.  
At 0-6~GPa, as mentioned in Section III.B, the 
Drude component is extremely narrow due to the 
large values of $\sigma(0)$.  
At 10~GPa, the mIR peak is not well resolved from 
the Drude component any more, but the fitting was 
nevertheless performed to evaluate the remaining 
mIR component, as shown in Fig.~6(b).  
Again, measured $\sigma(0)$,\cite{139-pressure} for 
example 4.5 $\times 10^4$~$\Omega^{-1}$cm$^{-1}$ at 
$P$=10~GPa and $T$=8~K, were used to reduce the 
uncertainty.  
The fitting results suggest that the 
SW of mIR peak at 10~GPa is still sizable, but 
is indeed much smaller than that at 6~GPa, as 
discussed in the main text.

The $E_{\rm mIR}$, FWHM, and SW of the mIR peak obtained 
from the fitting are displayed in Figs.~3(d) and 4(d) 
for CeRhIn$_5$ and YbNi$_3$Ga$_9$, respectively.  
Here, $E_{\rm mIR}$ is defined as the center of mass of 
the Lorentz peaks, namely the mean of $\omega_0$'s of 
the Lorentz oscillators weighted by their respective SW's.  
The SW of a Lorentz oscillator is defined as its area 
in $\sigma(\omega)$, and the SW of the mIR peak is 
the area of the fitted total mIR peak, namely the red 
solid curves in Fig.~6.  


\begin{thebibliography}{99}
\bibitem{medium3}
Glycerin has good characteristics as a 
pressure transmitting medium.\cite{medium1,medium2}  
It does not solidify at room $T$ up to about 5~GPa, 
and produces more hydrostatic pressure even at 
lower $T$ compared with solid media such as NaCl 
and KBr, which are commonly used for optical 
studies with DAC.  Therefore, we have used glycerin 
although using a liquid medium such as glycerin 
in a $R(\omega)$ study with DAC is technically 
much more challenging than using a solid medium.
\bibitem{comment5} 
Due to strong absorption by the diamond anvil, the 
spectral range between 0.23 and 0.3~eV could not 
be measured well.   
Therefore, the $R(\omega)$ spectra in this range 
were interpolated as indicated by the broken 
curves in Figs.~3(b) and 4(b).  
It is clear from Figs.~3(c) and 4(c) that these 
interpolations have only minor effects on the 
obtained $\sigma(\omega)$.
\bibitem{comment2} 
To make the plots, the following data have been used.  
$\gamma$=290~mJ/K$^2$mol for CeCoIn$_5$ at 0~GPa,\cite{basov} 
$\gamma_0$=10~mJ/K$^2$mol for LaCoIn$_5$,\cite{nakatsuji} 
$\gamma$=382~mJ/K$^2$mol for CeRhIn$_5$ at 2~GPa,\cite{fisher} 
and $\gamma_0$=5.7~mJ/K$^2$mol for LaRhIn$_5$.\cite{115-sc}
\bibitem{comment1} 
In Ref. \onlinecite{139-pressure}, only the magnetic 
part of the resistivity ($\rho_m$) at 
high $P$ is reported.   Therefore, we used the 
resistivity of LuNi$_3$Ga$_9$ at $P$=0 
(Ref.~\onlinecite{139-ohara}) as the 
non-magnetic part ($\rho_0$) to obtain 
the total resistivity $\rho=\rho_0 + \rho_m$, 
and the corresponding dc conductivity.   
In addition, $\rho_m$ data at 3~GPa are not 
given in Ref.~\onlinecite{139-pressure}, 
so we estimated $\rho_m$ at 3~GPa based on 
the data at $P$=0, 4, and 6~GPa given in 
Ref.~\onlinecite{139-pressure}.
\bibitem{comment3}
$\gamma$=80 and 230~mJ/K$^2$mol at $P$=3 and 6~GPa, 
respectively,\cite{139-gamma} were used.
\bibitem{comment9}
This is because the negatively charged surrounding of 
an Yb or Ce becomes closer to the $f$ electrons at high 
$P$, causing more Coulomb repulsion between the $f$ 
electron and the surrounding.    
Note that an Yb or Ce is positively charged locally, 
since it provides a few electrons into the spatially 
extended $c$ band.  
Hence the surrounding of an Yb or Ce should be 
negatively charged locally.
\end{thebibliography}
\end{document}